\documentclass[12pt]{article}
\usepackage{amsmath,amssymb,amsbsy,amsthm,latexsym,cite}

\title{The Interface of Noncommutative Geometry and Physics}

\author{Joseph C. V\'arilly%
  \thanks{Regular Associate of the Abdus Salam ICTP.\quad
          Email: \texttt{varilly@cariari.ucr.ac.cr}}\\[1pc]
Departamento de Matem\'atica, Universidad de Costa Rica,\\
2060 San Jos\'e, Costa Rica}

\advance\topmargin by -\headheight
\advance\topmargin by -\headsep
\evensidemargin=\oddsidemargin
\newcommand{\A}{\mathcal{A}}        
\renewcommand{\AA}{\mathbb{A}}      
\newcommand{\Ahat}{\Hat{A}}
\newcommand{\as}{\quad\text{as}\enspace} 
\newcommand{\C}{\mathbb{C}}         
\DeclareMathOperator{\ch}{ch}       
\newcommand{\Coo}{C^\infty}         
\newcommand{\D}{\mathcal{D}}        
\newcommand{\delslash}{\partial\mkern-9mu/} 
\newcommand{\Dl}{\Delta}            
\newcommand{\eps}{\varepsilon}      
\newcommand{\Ga}{\Gamma}            
\newcommand{\ga}{\gamma}            
\DeclareMathOperator{\GeV}{GeV}     
\renewcommand{\H}{\mathcal{H}}      
\renewcommand{\Hat}[1]{\widehat{#1}}  
\newcommand{\HH}{\mathbb{H}}        
\DeclareMathOperator{\Hom}{Hom}     
\DeclareMathOperator{\id}{id}       

\newcommand{\La}{\Lambda}           
\newcommand{\lab}{\boldsymbol{\lambda}} 
\newcommand{\Onda}[1]{\widetilde{#1}} 
\newcommand{\ox}{\otimes}           
\newcommand{\pd}[2]{\frac{\partial#1}{\partial#2}} 
\newcommand{\R}{\mathbb{R}}         
\DeclareMathOperator{\rank}{rank}   
\newcommand{\Rbar}{\overline{R}}    
\newcommand{\sepword}[1]{\quad\text{#1}\quad} 
\newcommand{\Sf}{\mathbb{S}}        
\newcommand{\stroke}{\mathbin\vert} 
\newcommand{\sul}{\mathfrak{su}}    
\newcommand{\T}{\mathbb{T}}         
\newcommand{\taub}{\boldsymbol{\tau}} 
\newcommand{\thalf}{\tfrac{1}{2}}   
\newcommand{\tihalf}{\tfrac{i}{2}}  
\def\top{\mathrm{top}}              
\DeclareMathOperator{\Tr}{Tr}       
\DeclareMathOperator{\tr}{tr}       
\newcommand{\x}{\times}             
\newcommand{\1}{\mathbf{1}}         
\newcommand{\7}{\dagger}            
\newcommand{\8}{\bullet}            
\renewcommand{\:}{\colon}           
\def\<#1,#2>{\langle#1\stroke#2\rangle} 
\def\(#1,#2){(#1\stroke#2)}

\newcommand{\sunset}{%
\parbox{15mm}{\begin{picture}(20,10)
\put(0,5){\line(1,0){10}}
\put(20,5){\circle{20}}
\put(10,5){\line(1,0){30}}
\end{picture}}
}
\newcommand{\iice}{%
\parbox{21mm}{\begin{picture}(20,10)
\put(0,5){\line(1,0){10}}
\put(20,5){\circle{20}}
\put(40.2,5){\circle{20}}
\put(30.2,5){\oval(40.5,40)[bl]}
\put(30.2,5){\oval(40,40)[br]}
\put(50.2,5){\line(1,0){10}}
\end{picture}}
}
\newcommand{\sweet}{%
\parbox{10mm}{\begin{picture}(20,10)
\put(0,5){\line(-1,2){5}}
\put(0,5){\line(-1,-2){5}}
\put(10,5){\circle{20}}
\put(20,5){\line(1,2){5}}
\put(20,5){\line(1,-2){5}}
\end{picture}}
}
\newcommand{\dsweet}{%
\parbox{15mm}{\begin{picture}(20,10)
\put(0,5){\line(-1,2){5}}
\put(0,5){\line(-1,-2){5}}
\put(10,5){\circle{20}}
\put(30.2,5){\circle{20}}
\put(40.2,5){\line(1,2){5}}
\put(40.2,5){\line(1,-2){5}}
\end{picture}}
}
\newcommand{\ragdoll}{%
\parbox{19mm}{\begin{picture}(20,10)
\put(0,5){\line(-1,2){5}}
\put(0,5){\line(-1,-2){5}}
\put(10,5){\circle{20}}
\put(20,5){\line(2,1){30}}
\put(20,5){\line(2,-1){30}}
\qbezier(35,5)(35,10)(40,15)
\qbezier(35,5)(35,0)(40,-5)
\qbezier(45,5)(45,10)(40,15)
\qbezier(45,5)(45,0)(40,-5)
\end{picture}}
}
\newcommand{\seeingeye}{%
\parbox{16mm}{\begin{picture}(20,10)
\put(10,5){\line(-2,-1){10}}
\put(10,5){\line(-2,1){10}}
\put(10,5){\line(2,1){30}}
\put(10,5){\line(2,-1){30}}
\qbezier(25,5)(25,10)(30,15)
\qbezier(25,5)(25,0)(30,-5)
\qbezier(35,5)(35,10)(30,15)
\qbezier(35,5)(35,0)(30,-5)
\end{picture}}
}

\makeatletter
\def\section{\@startsection{section}{1}{\z@}{-3.5ex plus -1ex minus
 -.2ex}{2.3ex plus .2ex}{\large\bf}}
\def\subsection{\@startsection{subsection}{2}{\z@}{-3.25ex plus -1ex
 minus -.2ex}{1.5ex plus .2ex}{\normalsize\bf}}
\makeatother

\begin{document}

\maketitle

\begin{abstract}
As a mathematical theory per se, noncommutative geometry (NCG) is by
now well established. From the beginning, its progress has been
crucially influenced by quantum physics: we briefly review this
development in recent years.

The Standard Model of fundamental interactions, with its central role
for the Dirac operator, led to several formulations culminating in the
concept of a real spectral triple. String theory then came into
contact with NCG, leading to an emphasis on Moyal-like algebras and
formulations of quantum field theory on noncommutative spaces. Hopf
algebras have yielded an unexpected link between the noncommutative
geometry of foliations and perturbative quantum field theory.

The quest for a suitable foundation of quantum gravity continues to
promote fruitful ideas, among them the spectral action principle and
the search for a better understanding of ``noncommutative spaces''.
\end{abstract}

\section{Introduction}

About 20 years ago, the mathematical theory nowadays known as
Noncommutative Geometry (NCG) began taking shape. A landmark paper of
Connes (1980) ushered in a differential geometric treatment of the
noncommutative torus~\cite{ConnesTorus} (further developed and
classified by Rieffel~\cite{RieffelRot}), which remains the paradigm
of a noncommutative space. Its differential calculus was put in a more
general framework at the Oberwolfach meeting in September--October
1981, where Connes unveiled a ``homology of currents for operator
algebras''~\cite{ConnesObW}, which soon became known as cyclic
cohomology~\cite{ConnesLambda}. This was developed in detail in his
``Noncommutative Differential Geometry''~\cite{ConnesNCDiffG}, in
preprint form around Christmas 1982; the related periodic cyclic
cohomology is a precise generalization, in algebraic language, of the
de~Rham homology of smooth manifolds.

The same algebraic approach, applied to the theory of
foliations~\cite{ConnesFoli}, led Connes to emphasize the notion of
Fredholm module, which is a cornerstone of his work with Karoubi on
canonical quantization~\cite{ConnesKaFred}. A key observation here is
that anomalous commutators form a cyclic 1-cocycle~\cite{ArakiSchw}, so
that in the noncommutative approach to quantum field theory, the
Schwinger terms are built~in.

Noncommutative geometry, then, is an operator-algebraic reformulation
of the foundations of geometry, extending to noncommutative spaces. It
allows consideration of ``singular spaces'', erasing the distinction
between the continuous and the discrete. On the mathematical side,
current topics of interest include index theory and groupoids,
mathematical quantization, the Baum--Connes conjectures on the
$K$-theory of group algebras, locally compact quantum groups, second
quantization in the framework of spectral triples, and the Riemann
hypothesis. Our focus here, however, is on its interface with physics.

\section{NCG and the Standard Model}

We say interface because one should not speak of the ``application''
of NCG to physics, but rather of mutual intercourse. Indeed, the first
use of noncommutative geometry in physics did not attempt to derive
the laws of physics from some NCG construct, but simply and humbly, to
learn from the mainstream physical theories ---concretely, the
Standard Model (SM) of fundamental interactions--- what the
(noncommutative) geometry of the world could be.

The crucial concepts of the SM are those of gauge fields and of chiral
fermions: they correspond to two basic notions of NCG, namely
connections and Dirac operators. Indeed, the algebraic definition of
linear connection is imported verbatim into NCG. Chiral fermions, for
their part, are acted on by Dirac and Dirac--Weyl operators.

Dirac operators are a source of NCG: any complex spinor bundle on a
smooth manifold $S \to M$ gives rise to a generalized Dirac operator
$D$ on the spinor space $L^2(M,S)$, whose
sign operator $F = D|D|^{-1}$ determines a Fredholm module; its
$K$-homology class $[F] \in K_\8(M)$ depends only on the underlying
spin$^c$-structure~\cite{BaumD,HigsonR}. Since the spin$^c$ structure
determines the orientation of the manifold, this fundamental class
---sometimes called a $K$-orientation~\cite{AtiyahBS}--- is a finer
invariant than the usual fundamental class in homology.

The approach to the SM by Connes and Lott~\cite{ConnesL} used a
noncommutative algebra to describe the electroweak sector, plus a
companion algebra to incorporate colour symmetries (see \cite{Sirius}
and \cite{KastlerS} for reviews of this preliminary approach). Later
on~\cite{ConnesReal}, a better understanding of the role of the charge
conjugation allowed this pair of algebras to be replaced by a
\textit{single} algebra acting bilaterally.

The gauge potentials appearing in the SM may be collected into a
single package of differential forms:
$$
\AA' = i(B,W,A),
$$
where
$$
B = -\tihalf g_1 \mathbf{B}_\mu \,dx^\mu, \quad
W = -\tihalf g_2\,\taub \cdot \mathbf{W}_\mu \,dx^\mu
\quad{\rm and}
\quad A = -\tihalf g_3\,\lab \cdot \mathbf{A}_\mu\,dx^\mu,
$$
with $\mathbf{B}$, $\mathbf{W}$ and $\mathbf{A}$ denoting respectively
the hypercharge, weak isospin and colour gauge potentials; $W$ is to
be regarded as a quaternion-valued $1$-form. Thus, $\AA'$ is an
element of $\La^1(M) \ox \A_F$, where the noncommutative algebra
$\A_F := \C \oplus \HH \oplus M_3(\C)$, that we have called the
``Eigenschaften algebra''~\cite{Cordelia}, plays the crucial role.

We next collect all chiral fermion fields into a multiplet $\Psi$ and
denote by $J$ the charge conjugation; then the fermion kinetic term is
rewritten as follows:
$$
I(\Psi,\AA',J) = \<\Psi, (i \delslash + \AA' + J\AA' J^\7) \Psi>.
$$
To incorporate the Yukawa part of the SM Lagrangian, let 
$\phi$ be a Higgs doublet with vacuum expectation value~$v/\sqrt2$, 
normalized by setting $\Phi := \sqrt2\,\phi/v$. We need both
$$
\Phi = \begin{pmatrix} \Phi_1\\ \Phi_2\end{pmatrix}
\sepword{and}
\Onda\Phi := \begin{pmatrix} -\bar\Phi_2 \\ \bar\Phi_1 \end{pmatrix}.
$$
The Higgs may be properly regarded as a quaternion-valued field; by
introducing $q_\Phi = \begin{pmatrix} \bar\Phi_1 & \bar\Phi_2 \\
- \Phi_2 & \Phi_1 \end{pmatrix}$, where
$\langle q_\Phi \rangle = 1$, we may write, schematically for a
right-left splitting of the fermion multiplets:
$$
\AA'' = \begin{pmatrix} & M^\7 (q_\Phi - 1) \\
(q_\Phi - 1)^\7 M & \end{pmatrix},
$$
where $M$ denotes the mass matrix for quarks (including the
Cabibbo--Kobayashi--Maskawa parameters) and leptons. Denoting by
$\D_F$ the Yukawa operator which relates the left- and right-handed
chiral sectors in the space of internal degrees of freedom, the Yukawa
terms for both particles and antiparticles (for the first generation)
can now be written as
\begin{align*}
I(\Psi,\AA'',J) &:= \<\Psi, (\D_F + \AA'' + J\AA''J^\7) \Psi>
\\
&= \bar q_L \Phi\,m_d\,d_R + \bar q_L \Onda\Phi\,m_u\,u_R
+ q_R \bar \Phi\,\bar m_d\, \bar d_L
+ q_R \Onda{\bar\Phi}\,\bar m_u\, \bar u_L
\\
&\qquad
+ \bar\ell_L \Phi\,m_e\,e_R
+ \bar\ell_L \Onda\Phi\,m_\nu\,\nu_R
+ \ell_R \bar\Phi\, \bar m_e\, \bar e_L 
+ \ell_R  \Onda{\bar\Phi}\,\bar m_\nu\, \bar \nu_L
+ {\rm h.c.}
\end{align*}
Altogether, we get a Dirac--Yukawa operator
$\D = i\delslash \oplus \D_F$. With $\AA := \AA'\,\oplus\,\AA''$, the
\textit{whole} fermionic sector of the SM is recast as
$$
I(\Psi,\AA,J) = \<\Psi, (\D + \AA + J\AA J^\7) \Psi>.
$$

The upshot is that the ordinary gauge fields and the Higgs are
combined as entries of a \textit{generalized gauge potential}. The
Yukawa terms come from the minimal coupling recipe applied to the
gauge field in the internal space. The Dirac--Yukawa operator is seen
to contain in NCG all the relevant information pertaining to the~SM.

This Connes--Lott reconstruction of the SM gave rise to two
``predictions''. (At that time, the top quark had not yet been seen,
and the best estimates for its mass ranged around $130 \GeV$.) The NCG
model sort of explains why the masses of the top quark, the $W$ and
$Z$ particles and the Higgs particle should be of the same order, and
gave right away
$$
m_\top \geq \sqrt{3}\, m_W \approx 139 \GeV.
$$
With a bit of renormalization group running~\cite{Orpheus}, it fell
right on the mark. On the other hand, the ``prediction'' for the Higgs
mass from Connes' NCG has remained stuck around $200 \GeV$, while the
current phenomenological prejudice is that it should be much lower.

A major limitation of the Connes--Lott approach is that the fermion
mass matrix must be taken as an input. A different though less
ambitious proposal, put forward about the same time, was the
Mainz--Marseille scheme, based on organizing the $(W,B)$ forms and the
Higgs field components as a $3 \x 3$ matrix in the Lie superalgebra
$\sul(2|1)$. The known families of quarks and leptons can then be
fitted into (reducible but indecomposable) $\sul(2|1)$
representations, and some relations among the quark masses and CKM mixing
parameters emerge~\cite{Scheck}; this analysis applies likewise to 
lepton masses and neutrino mixing.

\vspace{6pt}

This ``bottom-up'' interaction between physics and NCG yielded an
important dividend. The clarification of the role of $J$ as a ``Tomita
conjugation'' \cite{Takesaki} ostered the emergence of the concept of
a \textit{real spectral triple} ---the word ``real'' being taken in
the sense of Atiyah's ``Real $K$-theory''~\cite{AtiyahReal}--- which
led to a construction of noncommutative spin
manifolds~\cite{ConnesCours}. This construction explained in our
\textit{Elements of Noncommutative Geometry}~\cite{Polaris}. Thus, we
now know how to put fermion fields on a noncommutative manifold.

A \textit{spectral triple} $(\A,\H,D)$ consists of a (unital)
\textit{algebra} $\A$ represented on a Hilbert space $\H$, plus a
selfadjoint operator $D$ on~$\H$, such that $[D,a]$ is bounded for all
$a \in \A$ and $D^{-1}$ is compact. It is \textit{even} if there is a 
grading operator $\chi$ (or ``$\ga_5$'') on~$H$ wrt which $\A$ is even 
and $D$ is odd. It is \textit{real} if there is an antiunitary operator
$J$ on~$H$ such that $J^2 = \pm 1$, $JD = \pm DJ$ and
$J\chi = \pm\chi J$ (even case); the signs depend on a certain 
dimension $\bmod 8$. From these data, by imposing a few extra
conditions, spin manifolds can be reconstructed~\cite{ConnesGrav}.

\section{The spectral action principle}

The early Connes--Lott models did not take account of gravity. To
remedy that, Connes and Chamseddine~\cite{ChamseddineCSpec} proposed a
universal formula for an action associated with a noncommutative spin
geometry, modelled by a real spectral triple $(\A,\H,D,J)$. The action
$S(D) = B_\phi[D] + \<\Psi, D\Psi>$ is based on the spectrum of the Dirac
operator and is a geometric invariant. Automorphisms of the algebra
$\A$ combine ordinary diffeomorphisms with internal symmetries which
alter the metric by $D \mapsto D + A + J A J^\7$.

The bosonic part of the action functional is
$B_\phi[D] = \Tr \phi(D^2)$, where $\phi$ is an ``arbitrary'' positive
function (a regularized cutoff) of~$D$. Chamseddine and Connes argue
that $B_\phi$ has an asymptotic expansion 
$$
B_\phi[D/\La] \sim \sum_{n=0}^\infty f_n\, \La^{4-2n}\, a_n(D^2)
\as \La \to \infty,
$$
where the $a_n$ are the coefficients of the heat kernel expansion for
$D^2$ and $f_0 = \int_0^\infty x \phi(x) \,dx$,
$f_1 = \int_0^\infty \phi(x) \,dx$, $f_2 = \phi(0)$,
$f_3 = -\phi'(0)$, and so on: this is in fact a Ces\`aro asymptotic
development~\cite{Odysseus}. On computing this expansion for the
Dirac--Yukawa operator of the Standard Model, they found all terms in
the bosonic part of the SM action, plus unavoidable gravity couplings.
That is to say, the spectral action for the Standard Model unifies
with gravity at a very high energy scale.

Recently, Wulkenhaar~\cite{WulkenhaarTheta} has conjectured that on
$\theta$-deformed spacetime, the spectral action may have the
necessary additional symmetries to renormalize gauge theories. In this
regard, Langmann~\cite{LangmannAction} has managed to prove that the
effective action of fermions coupled to a Yang--Mills field contains
the usual Yang--Mills bosonic action. To check the conjecture, one
first needs to extend the spectral action to the context of noncompact
NC manifolds.

By ``noncompact noncommutative spin geometry'' we understand a real
spectral triple $(\A,\H,D,J)$ where $\A$ is a \textit{nonunital}
algebra, where $[D,a]$ is bounded and $a|D|^{-1}$ is compact for all
$a \in \A$. Geometries of this type are discussed in 
\cite{Selene,RennieSwan,RennieSmooth}: the analytic toolbox of NC spin 
geometries~\cite{Polaris} extends to the noncompact case if suitable 
multiplier algebras are employed.

\section{Noncommutative field theory}

The next phase of the dialogue between NCG and the physics of
fundamental interactions was characterized by a ``top-down'' approach.
An important precursor is the 1947 paper by Snyder, ``Quantized
space-time''~\cite{Snyder}, where it was first suggested that
coordinates $x^\mu$ may be noncommuting operators; the six commutators
are of the form $[x^\mu,x^\nu] = (ia^2/\hbar)\,L^{\mu\nu}$ where $a$
is a basic unit of length and the $L^{\mu\nu}$ are generators of the
Lorentz group; throughout, Lorentz covariance is maintained. Then as
now, noncommuting coordinates were used to describe spacetime in the
hope of improving the renormalizability of QFT and of coming to terms
with the nonlocality of physics at the Planck scale.

In a similar vein, Doplicher, Fredenhagen and
Roberts~\cite{DoplicherFR} have considered a model with commutation
relations
$$
[x^\mu, x^\nu] = i\, Q^{\mu\nu},
$$
where the $Q^{\mu\nu}$ are the components of a tensor, but commute
among themselves and with each~$x^\mu$. Thus in their formalism,
Lorentz invariance is also explicitly kept.

String theorists have recently revived this top-down approach. In
their most popular model, the commutation relations are simply of the
form
\begin{equation}
[x^\mu, x^\nu] = i\, \theta^{\mu\nu},
\label{eq:comm-reln}
\end{equation}
where the $\theta^{\mu\nu}$ are $c$-numbers, breaking Lorentz
invariance. As anticipated by Sheikh-Jabbari \cite{Jabbari} and
plausibly argued by Seiberg and Witten~\cite{SeibergWGeom}, open
strings with allowed endpoints on 2D-branes in a B-field background
act as electric dipoles of the abelian gauge field of the brane; the
endpoints live on the noncommutative space determined
by~\eqref{eq:comm-reln}, as pointed out by~\cite{Schomerus}.

Slightly before, Connes, Douglas and Schwarz~\cite{ConnesDS} had shown
that compactification of $M$-theory, in the context of dimensionally
reduced gauge theory actions, leads to spaces with embedded
noncommutative tori. See also~\cite{LandiLS} for the relation between
noncommutative geometry and strings.

An important feature of~\cite{SeibergWGeom} is the ``Seiberg--Witten
map'' in gauge theory, which relates gauge fields and gauge variations
in a noncommutative theory with their commutative counterparts. In the
NC theory, multiplication is replaced by the Moyal product
$\star_\theta$ with parameter $\theta = [\theta^{\mu\nu}]$; in order
to preserve gauge equivalence (whenever $A$ and $A'$ are equivalent
gauge fields, so should be the NC gauge fields $\Ahat$ and $\Ahat'$),
Seiberg and Witten found $\theta$-dependent formulas for the latter.
As explained by Jackiw and Pi~\cite{JackiwP} (see
also~\cite{JurcoSW}), these formulas correspond to an infinitesimal
1-cocycle for a projective representation of the underlying gauge
group in the Moyal algebra.

The Moyal product which appears here is nonperturbatively defined, for
nondegenerate skewsymmetric $\theta$, as
$$
f \star_\theta g(u) := (\pi\theta)^{-4}
\int_{\R^4\x\R^4} f(u + s)g(u + t)\,e^{2is\theta^{-1}t}\,d^4s\,d^4t,
$$
and this gives rise to the commutation relations~\eqref{eq:comm-reln}.
Those are just the commutation relations of quantum mechanics, when
$\hbar$ replaces~$\theta$! The precise relation of this integral
formula to the asymptotic development usually put forward as the
Moyal product was spelled out some time ago in~\cite{Nereid}. This
product is the basis of the Weyl--Wigner--Moyal or phase-space
approach to quantum mechanics~\cite{Moyal}, which already had a long
history when (a version of) the Moyal product was rediscovered by
string theorists. It should be said that many of the recent papers
which purport to use this product in string theory or NC field theory
are rather careless; some are unaware of the mathematical properties
of the Moyal product, which are outlined, for instance, in our
\cite{Phobos,Deimos} or in~\cite{Selene}.

It is worth pointing out that noncommutative field theory can be
developed independently of its string theory motivation, and indeed
preexisted the Seiberg--Witten paper. Quantum field theory has an
algebraic core which is independent of the nature of spacetime. From
the representation theory of the infinite dimensional orthogonal group
(or an appropriate subgroup), with the input of a one-particle space,
one can derive all Fock space quantities of interest: nothing really
changes if the ``matter field'' evolves on a noncommutative space.
That is to say, one can apply the canonical quantization machinery to
a noncommutative kind of one-particle space~\cite{Atlas}. The
long-standing hope, that giving up \textit{locality} in the
interaction of fields would be rewarded with a better ultraviolet
behaviour, was now amenable to rigorous scrutiny, and it is not borne
out. QFT on noncommutative manifolds also requires renormalization.
This, in some sense the first result of NCFT, was proved in general by
Gracia-Bond\'{\i}a and myself in~\cite{Atlas}, using a cohomological 
argument internal to noncommutative geometry.

Of course, one can prove the same in the context of a
\textit{particular} NCG model, by writing down the integral
corresponding to a Feynman diagram, and finding it to be divergent.
That had been shown previously by Filk~\cite{Filk}, for the scalar
Lagrangian theory associated to the Moyal product algebra. Filk made
the point that the momentum integrals for planar Feynman graphs are
identical to those in the commutative theory, and the contributions
from nonplanar graphs cannot cancel them. The same basic point had
been made much earlier in~\cite{GonzalezAKA}, with regard to the
continuum limit of a reduced model of large~$N$ field theory.

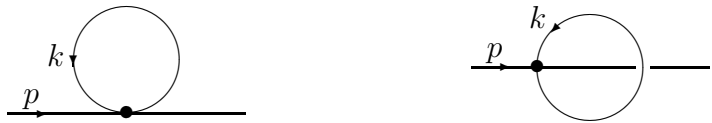
\begin{figure}[ht]
\centering
\vspace{1pc}
\parbox{50pt}{\begin{picture}(40,40)
\put(-10,5){\line(1,0){90}}
\put(-5,5){\vector(1,0){10}}
\put(-4,9){$p$}
\put(32,2.5){$\bullet$}
\put(35,25.5){\circle{40}}
\put(14.8,24){\vector(0,-1){1}}
\put(5,23){$k$}
\end{picture}}
\hspace{10em}
\parbox{50pt}{\begin{picture}(40,5)
\put(-10,5){\line(1,0){62}}
\put(-5,5){\vector(1,0){10}}
\put(-4,9){$p$}
\put(12,2.5){$\bullet$}
\put(35,5){\circle{40}}
\put(21,19){\vector(-1,-1){1}}
\put(12,20){$k$}
\put(58,5){\line(1,0){22}}
\end{picture}}
\vspace{1pc}
\caption{Planar and nonplanar tadpole diagrams in NC $\phi^4$ theory}
\label{fig:nonplanar}
\end{figure}

The distinction between planar and nonplanar Feynman diagrams is an
essential feature of NC field theory. Consider, for instance, the
theory given by the action functional
$$
S = \int d^4x\, \biggl(\frac{1}{2} \pd{\phi}{x^\mu} \pd{\phi}{x_\mu} 
+ \frac{1}{2} m^2\phi^2 + \frac{g}{4!}
  \phi \star_\theta \phi \star_\theta \phi \star_\theta \phi \biggr).
$$
The Feynman rules yield the same propagators as in the commutative
theory, but the vertices get in momentum space an extra
factor proportional to
$$
\exp\biggl(-\frac{i}{2}
\sum_{k<l} p_{k\mu} \theta^{\mu\nu} p_{l\nu} \biggr),
$$
where $p_1,\dots,p_r$ are the momenta incoming on the vertex, in
cyclic order. Planar diagrams get overall phase factors depending only
on external momenta; for nonplanar diagrams, the phase factors may
also depend on loop variables, and the corresponding integrals may
become convergent. For the tadpole diagrams of
Figure~\ref{fig:nonplanar}, we get amplitudes of the form
$$
\Ga_{\rm pl}(p)\propto \int \frac{d^4k}{k^2 + m^2},  \qquad
\Ga_{\rm npl}(p)\propto \int \frac{d^4k}{k^2 + m^2}\, e^{-ip\theta k},
$$
and the second integral is finite for $p \neq 0$.

However, nonplanar diagrams may become divergent again for particular
values of the momenta (try $p = 0$ in the previous example). For
complicated diagrams with subdivergences, this dependence of the
amplitude behaviour on~$p$ is troublesome, because such diagrams may
unexpectedly become divergent again. This is the notorious UV/IR
mixing~\cite{MinwallaRS}, which tends to spoil renormalizability. For
Moyal NC Yang--Mills theory, this happens already at the 2-loop level.

It was pointed out by Gomis and Mehen~\cite{GomisM} that whenever
there is timelike noncommutativity $\theta^{0i} \neq 0$, one
encounters a violation of unitarity of the $S$-matrix. However, Bahns
et~al~\cite{BahnsDFP} have argued that, if the above Lagrangian
approach is replaced by a Hamiltonian approach to NC field theory, the
apparent failure of unitarity disappears. The change in viewpoint
concerns only the nonplanar diagrams. This should not really be
surprising: even for ordinary Yang--Mills theories, the full
equivalence of both approaches has never been
established~\cite{Cheng}. Actually, the indications seem to be that
they are \textit{not} equivalent in noncommutative field
theory~\cite{RimY,LiaoS}.

The literature on NC field theories is already very large, and of
uneven quality; we cannot really do it justice here. For recent
extensive reviews, see \cite{DouglasN} and \cite{SzaboReview}.

Finally, it is now possible to combine these NC field theories with
the Connes--Lott approach, by taking the tensor product $\A_{NC} \ox
\A_F$ of a noncommutative spacetime algebra and the Eigenschaften
algebra $\A_F$. This has been done by Morita~\cite{Morita}, Chaichian
et~al \cite{ChaichianPJTModel} and the M\"unchen
group~\cite{CalmetJSWW}, in various ways (and with different
outcomes). However, there is some doubt as to whether these models are
anomalous~\cite{Camilla}, and therefore nonrenormalizable~\cite{Nysa}.

\section{Noncommutative spaces}

In an eventual noncommutative approach to quantum gravity, one must be
able to sum over families of noncommutative spaces. This, together
with the need for good examples, has inspired a search for
noncommutative manifolds. Connes~\cite{ConnesSurvey} has suggested
that ``NC spheres'' may be obtained from two homological conditions:
(1) the Chern character form vanishes in all intermediate degrees; and
(2) the metric may vary while keeping the volume form fixed. In even
dimensions $n = 2m$, this comes down to setting
\begin{subequations}
\label{eq:ch-eqns}
\begin{gather}
\ch_k(e) \equiv (-1)^k \frac{(2k)!}{k!} \tr((e - \thalf)\,(de\,de)^k)
= 0 \sepword{for} k = 0,1,\dots,m-1,
\label{eq:ch-vanish}  \\
\pi_D(\ch_m(e)) = \chi,
\label{eq:ch-volume}
\end{gather}
\end{subequations}
where $e = e^2 = e^* \in M_{2^m}(\A)$ is an orthogonal projector,
$\chi$ is the chiral grading operator on~$\H$, and
$\pi_D(a_0\,da_1\dots da_n) := a_0\,[D,a_1]\dots [D,a_n]$; the domain
of~$\pi_D$ is the universal graded differential algebra over~$\A$,
which may be regarded as the space of chains for Hochschild homology.
Briefly, one finds that \eqref{eq:ch-vanish} makes $\ch_m(e)$ a
Hochschild cycle, and the \eqref{eq:ch-volume} says that this cycle
gives the desired volume form~\cite{ConnesGrav}. Now
\eqref{eq:ch-eqns} becomes a system of equations which impose severe
restrictions on the algebra $\A$ to which the matrix elements of~$e$
belong.

For $n = 2$, there is only the commutative
solution~\cite{ConnesSurvey}, $\A = \Coo(\Sf^2)$. For $n = 4$, Connes
and Landi~\cite{ConnesLa} found ``$\theta$-twisted 4-spheres''
$\Sf^4_\theta$ with embedded copies of the NC 2-torus $\T^2_\theta$.
Later, Connes and Dubois-Violette~\cite{ConnesDV} showed that there is
a 3-parameter family of NC 3-spheres, including a $\theta$-twisted
subfamily $\Sf^3_\theta$. These $\theta$-twisted spheres can all be
described as quantum homogeneous spaces~\cite{ConnesDV,Larissa}: in
fact, we can construct $M_\theta$ by twisting whenever $M = G/H$ is a
quotient of compact Lie groups with $\rank H \geq 2$. The
noncommutative algebra $\Coo(M_\theta)$ is simply $\Coo(M)$ equipped
with a periodic version of the Moyal product~\cite{RieffelDefQ}, and
the symmetry group $G$ is correspondingly deformed to a quantum
group~\cite{Larissa}. If $M$ is spin, then so is $M_\theta$; it
carries a NC spin geometry obtained by isospectral deformation from
that of~$M$~\cite{ConnesLa}. Noncommutative twistors can also be
obtained in this manner.

One motivation for constructing such examples of noncommutative spaces
is to come back to quantum gravity by (a) allowing for metric
fluctuations with fixed volume; and eventually (b) relaxing the
Hochschild condition to incorporate ``virtual'' NC manifolds, whereby
the condition \eqref{eq:ch-volume} would appear as the signal of a
``true'' manifold~\cite{ConnesTalkOW}.

\section{The Connes--Kreimer Hopf Algebras}

Bogoliubov's renormalization scheme in dimensional regularization can
be summarized as follows. Let $\Ga$ be a one-particle irreducible
(1PI) graph (i.e., a connected graph which cannot be disconnected by
removing a single line), with amplitude $f(\Ga)$; if $\Ga$ is
primitive (i.e., has no subdivergences), set
$$
C(\Ga) := - T(f(\Ga)),  \sepword{and then}  R(\Ga) := f(\Ga) + C(\Ga),
$$
where $C(\Ga)$ is the \textit{counterterm}, $R(\Ga)$ is the desired
finite value, and $T$ projects on the pole part: in other words, for
primitive graphs, one simply removes the pole part. We recursively
define Bogoliubov's $\Rbar$-operation by setting
$$
\Rbar(\Ga) = f(\Ga)
 + \sum_{\emptyset\subsetneq\ga\subsetneq\Ga} C(\ga)\,f(\Ga/\ga),
$$
where $C(\ga_1\dots\ga_r) := C(\ga_1) \dots C(\ga_r)$, whenever
$\ga = \ga_1\dots\ga_r$ is a disjoint union of several pieces. Then we
remove the pole part of the previous expression:
$C(\Ga) := - T(\Rbar(\Ga))$ and $R(\Ga) := \Rbar(\Ga) + C(\Ga)$. 
Overall,
\begin{subequations}
\label{eq:graph-renorm}
\begin{align}
C(\Ga) &:= - T\biggl[ f(\Ga)
+ \sum_{\emptyset\subsetneq\ga\subsetneq\Ga}C(\ga)\,f(\Ga/\ga)\biggr],
\label{eq:graph-renorm-C}
\\
R(\Ga) &:= f(\Ga) + C(\Ga)
 + \sum_{\emptyset\subsetneq\ga\subsetneq\Ga} C(\ga)\,f(\Ga/\ga).
\label{eq:graph-renorm-R}
\end{align}
\end{subequations}

Now let $\Phi$ stand for any particular QFT. There is an associated
Hopf algebra $H_\Phi$ \cite{ConnesKrRHI,ConnesKrRHII} which is, first
of all, a commutative algebra generated by the 1PI graphs~$\Ga$
of~$\Phi$. The product is given by the disjoint union of graphs. The
counit $\eps$ is defined on generators by $\eps(\Ga) := 0$ unless
$\Ga$ is empty, and $\eps(\emptyset) := 1$; and the unit map $\eta$ is
determined by $\eta(1) := \emptyset$. The \textit{coproduct} $\Dl$ is
given by
$$
\Dl\Ga := \sum_{\emptyset\subseteq\ga\subseteq\Ga} \ga \ox \Ga/\ga,
$$
where the sum ranges over all subgraphs $\ga$ which are divergent and
proper (i.e., removing one internal line cannot make more connected
components); $\ga$ itself need not be connected. The terms for
$\ga = \emptyset$ and $\ga = \Ga$ in the sum are
$\Ga \ox 1 + 1 \ox \Ga$. The notation $\Ga/\ga$ denotes the
(connected, 1PI) graph obtained from~$\Ga$ by replacing each component
of $\ga$ by a single vertex. One checks that $\Dl$ is
coassociative~\cite{ConnesKrRHI}, so $H_\Phi$ is a bialgebra.

Here are some coproducts for $\Phi = \varphi_4^4$, taken 
from~\cite{Etoile}:

\begin{figure}[ht]
\centering
$$
\Dl \Bigl( \sunset \Bigr) = \1 \ox \sunset + \sunset \ox \1
$$
\caption{The ``setting sun'': a primitive diagram$^1$}
\label{fig:sunset}
\end{figure}

\stepcounter{footnote}
\footnotetext{Note: the setting sun diagram is primitive in position
space, and is usually not considered primitive when working in
momentum space. Of course, the associated amplitude must not depend on
this description; and it does not~\cite{Zimmermann}. This piece of
wisdom seems less well known than it should be.}

\begin{figure}[ht]
\centering
$$
\Dl \Bigl( \iice \Bigr) = 
\1 \ox \iice + 2\ \sweet \ox \sunset \put(-35,0){$\bullet$}
+ \iice \ox \1
$$
\caption{The ``double ice cream in a cup'' (depth = 2)}
\label{fig:icecream}
\end{figure}

\begin{figure}[ht]
\centering
$$
\Dl \Bigl(\,\,\, \ragdoll \Bigr) = \1 \ox \,\ragdoll + \
\sweet \ox \seeingeye\put(-39,0){$\bullet$} + \
\sweet \ox \,\dsweet\put(-6,0){$\bullet$}
$$
$$
+ \ \seeingeye \ox \,\sweet\put(-12,0){$\bullet$} + \
\sweet\ \sweet \ox \,\sweet\put(-32,0){$\bullet$}
\put(-12,0){$\bullet$} + \ \ragdoll \ox \1
$$
\caption{The ``rag-doll'' (depth = 3)}
\label{fig:ragdoll}
\end{figure}

A grading, which ensures that $H_\Phi$ is a Hopf algebra, is provided
by \textit{depth}~\cite{Nysa}: a graph $\Ga$ has depth $k$ (or is
``$k$-primitive'') if
$$
P^{\otimes(k+1)}(\Dl^k\Ga) = 0  \sepword{and}
P^{\otimes k}(\Dl^{k-1}\Ga) \neq 0.
$$
where $P$ is the projection $\eta\eps - \id$. Depth measures the
maximal length of the inclusion chains of subgraphs appearing in the
Bogoliubov recursion. In dimensional regularization, a graph of
depth~$l$ is expected to display a pole of order~$l$. The antipode $S$ 
can now be defined as the inverse of $\id = \eta\eps - P$
for the convolution; if $\Ga_l$ is a graph of depth $l$, one finds
\begin{equation}
S(\Ga_l) := \sum_{k=1}^l P^{*k}\,\Ga_l = - \Ga_l
+ \sum_{\emptyset\subsetneq\ga\subsetneq\Ga_l} S(\ga)\,\Ga_l/\ga.
\label{eq:graph-antp}
\end{equation}

As it stands, the Hopf algebra $H_\Phi$ corresponds to a formal
manipulation of graphs. These formulas can be matched to expressions
for numerical values, as follows. The Feynman rules for the
unrenormalized theory prescribe an algebra homomorphism
$$
f : H_\Phi \to \A
$$
into some commutative algebra $\A$; that is, $f$ is linear and
$f(\Ga_1 \Ga_2) = f(\Ga_1)\,f(\Ga_2)$. In dimensional regularization,
$\A$ is an algebra of Laurent series in a complex parameter~$\eps$,
and $\A$ is the direct sum of two \textit{subalgebras}:
$$
\A = \A_+ \oplus \A_-,
$$
where $\A_+$ is the holomorphic subalgebra of Taylor series and $\A_-$
is the subalgebra of polynomials in $1/\eps$ without constant term.
The projection $T\: \A \to \A_-$, with $\ker T = \A_+$, picks out the
pole part, in a minimal subtraction scheme. Now $T$ is not a
homomorphism, but the property that both its kernel and image are
subalgebras is reflected in a ``multiplicativity constraint'':
$$
T(ab) + T(a)\,T(b) = T(T(a)\,b) + T(a\,T(b)).
$$

The equation \eqref{eq:graph-renorm-C} means that ``the antipode
delivers the counterterm'': one replaces $S$ in the calculation
\eqref{eq:graph-antp} by~$C$ to obtain the right hand side, before
projection with~$T$. From the definition of the coproduct in $H_\Phi$,
\eqref{eq:graph-renorm-R}, which extracts the finite value, is a
\textit{convolution} in $\Hom(H_\Phi,\A)$, namely, $R = C * f$. To
show that $R$ is multiplicative, it is enough to verify that the
counterterm map $C$ is multiplicative: convolution of homomorphisms is
a homomorphism because $\A$ is commutative. The multiplicativity
of~$C$ follows from the constraint on~$T$, as shown by Connes and
Kreimer in~\cite{ConnesKrRHI}. See also \cite{Calypso} and
\cite{GirelliMK} in regard to these convolution formulas.

The previous discussion is logically independent of NCG, but there is
an important historical link. The Hopf algebra approach to
renormalization theory arose in parallel with the Connes--Moscovici
noncommutative theory of foliations. In that theory, a foliation is
described by a noncommutative algebra of functions twisted by local
diffeomorphisms, $\A = \Coo_c(F) \rtimes \Ga$; horizontal and vertical
vector fields on the frame bundle $F \to M$ are represented on $\A$ by
the action of a certain Hopf algebra $H_{CM}$ which provides a way to
compute a local index formula in NCG~\cite{ConnesMHopf}. One can map
$H_{CM}$ into an extension of the \textit{Hopf algebra of rooted
trees}, a precursor of the Connes--Kreimer graphical Hopf algebras
which is described in detail in~\cite{ConnesKrHopf}
and~\cite{Polaris}. On extending the Hopf algebra $H_\Phi$ of graphs
by incorporating operations of insertion of subgraphs, one obtains a
noncommutative Hopf algebra of the $H_{CM}$ type, which gives a
supplementary handle on the combinatorial structure of
$H_\Phi$~\cite{ConnesKrLie}.

\section{Outlook}

Noncommutative Geometry has had, for many years now, a mutually
rewarding conversation with quantum physics. The underlying motif of
this conversation can be said to be the belief that Quantum Field
Theory encodes the true geometry of the world, and that the
mathematical task is to elucidate this geometrical structure. The
payback to physics takes the form of new tools and methods; and the
work is far from over. For the biggest challenge, that of
understanding quantum gravity, there is a long way yet to travel.

Just as the effort to understand the gauge symmetries of the Standard
Model led, in due time, to the introduction of real structures for
spectral triples and from there to a noncommutative understanding of
spin geometries, we may likewise expect that the NC approach to
gravity will help to clarify our still imperfect understanding of the
nature of noncommutative manifolds. The story continues~\dots

\subsection*{Acknowledgements}

I am grateful to the organizers of the 6$^{\mathrm{th}}$~Conference on
Clifford Algebras and their Applications in Mathematical Physics for
inviting me to speak about noncommutative geometry and its interface
with physics. Support from the organizers and the Vicerrector\'{\i}a
de Investigaci\'on of the Universidad de Costa Rica is acknowledged.
Much of what is written here emerged from lengthy discussions and past
collaborations with Jos\'e M. Gracia-Bond\'{\i}a. Finally, I owe a
debt of gratitude to Alain Connes, whose continuing creation of
noncommutative geometry is an inspiration to us all.

\end{document}